\documentclass[aps,prd,preprint,amsmath,amssymb,nofootinbib,preprintnumbers]{revtex4-1}

\usepackage{epsf}
\usepackage{natbib}

\usepackage{graphicx}% Include figure files
\usepackage{dcolumn}% Align table columns on decimal point
\usepackage{bm}% bold math
\usepackage{amsmath}
\usepackage{amsfonts}
\usepackage{subfigure}

\def\be{\begin{equation}}
\def\beq\begin{equation}
\def\ee{\end{equation}}
\def\bea{\begin{eqnarray}}
\def\eea{\end{eqnarray}}

\def\beq{\begin{equation}}
\def\eeq{\end{equation}}
\def\beqa{\begin{eqnarray}}
\def\eeqa{\end{eqnarray}}

\newcommand{\bmat}{\left(\begin{array}}
\newcommand{\emat}{\end{array}\right)}

\newcommand{\half}{\frac{1}{2}}
\newcommand{\inv}[1]{\frac{1}{#1}}

\newcommand{\bit}{\begin{itemize}}
\newcommand{\eit}{\end{itemize}}
\newcommand{\bnu}{\begin{enumerate}}
\newcommand{\enu}{\end{enumerate}}
\newcommand{\deth}{\sqrt{h}}
\newcommand{\ba}{\begin{align}}

\begin{document}

\title{Hamiltonian perturbation theory in $f(R)$ gravity}

\author{Tuomas Multam\"aki}
\author{Jaakko Vainio}
\author{Iiro Vilja}

\affiliation{Department of Physics and Astronomy, University of Turku, 
FIN-20014, Finland}

%\date{May 3, 2006}

\begin{abstract}
Hamiltonian perturbation theory is used to analyse the stability of
$f(R)$ models. The Hamiltonian equations for the metric and its momentum conjugate are written for $f(R)$
Lagrangian in the presence of perfect fluid matter. The perturbations examined 
are perpendicular to $R$. As perturbations are added to the metric and momentum 
conjugate to the induced metric instabilities are found, depending on the form of
$f(R)$. Thus the examination of these instabilities is a way to rule out 
certain $f(R)$ models.
\end{abstract}

\maketitle

\section{Introduction}

The question of dark energy has been at the heart of 
cosmology since the discovery of accelerating expansion of 
the universe \cite{riess98}. The traditional picture of general relativity 
with ordinary relativistic or non-relativistic matter in homogeneous
and isotropic universe meets severe problems when accommodating it to 
current cosmological 
observations. The conflicting observational evidence 
comes mainly from supernova light curves \cite{riess98,perlmutter99}, 
CMB anisotropies \cite{Bennett2003,Netterfield2001} and large 
scale structures \cite{Perlmutter1999,Bunn1996}. This has 
lead to several suggested remedies. Perhaps the most popular 
way is to add some non-conventional matter to the universe. Among these
the simplest possibility is no doubt to use the cosmological constant. 
A review of the subject can be found in \cite{peebles2002}. In any case, 
the key aspect is the negative pressure of the new matter which boosts
the expansion of the universe. Other considerations include more
general distribution of 
matter, {\it i.e.} non-homogeneous or non-isotropic universe (see {\it e.g.} 
\cite{Alnes2005}). 

Besides these two, a lot of effort has been put into studies on  
generalizations and modifications of General Relativity. 
For example metric-affine theories (see {\it e.g.} \cite{Sotiriou2006}), 
scalar-tensor theory 
(see {\it e.g.} \cite{faraoni04,Bransdicke1961}), brane-world gravity 
(see {\it e.g.} \cite{Maartens2003}) and more general 
Lagrangians have been considered. In the present paper we are especially 
interested in $f(R)$ gravity models in which the 
Einstein-Hilbert action is replaced by a function of the 
curvature scalar $R$ \cite{Carroll2003,Carroll2004,Meng2003,
Allemandi2004,Sotiriou2008}. None of these modifications is 
free of problems and this is indeed the case of $f(R)$ gravity 
as well. As for any model, the cosmological observations issue some 
constraints (see {\it e.g.} \cite{Tsujikawa2007,Starobinsky2007}) as do the 
observations in the solar system (see {\it e.g.} \cite{Chiba2003,
Hu2007,Magnano93,multamaki2008,Henttunen2007}). The opinions are still 
divided on the viability of $f(R)$ theories of gravity. There 
are numerous approving studies (see {\it e.g.} \cite{Faraoni2006,Olmo2006}) 
as well as sceptical ones (see {\it e.g.} \cite{Faulkner2006,Erickcek2006}).

As the actual universe is not homogeneous and isotropic but
contains local perturbations, additional challenges for $f(R)$ 
theories emerge from stability analysis 
\cite{Dolgov2003,Soussa2003,Faraoni2005}. An acceptable cosmological
model has to be stable against perturbations in the metric and the mass
distribution. However, stability analysis is customarily done only in the 
direction of $R$, {\it i.e.} only curvature perturbations are considered. This is motivated in particular in the case of General 
Relativity, where the relation between space-time curvature and the
matter density is a simple one: the trace of Einstein equations imply
$R\propto \rho$. This in turn implicates a 
simple and direct relation between the perturbations in matter and curvature. This is not the only possibility. 
In a $f(R)$ model the model the relation is more complicated due to
appearance of function $f(R)$ and higher derivative terms in the field equations. 
The phase space is considerably larger and metrics 
corresponding a given matter distribution ambiguous. The physical
acceptability, however, of a model requires general stability; also
stability against perturbations which keep curvature constant,
perpendicular to $R$.

The Hamiltonian formulation of general relativity has been around since 
the work of Arnowitt, Deser and Misner \cite{Arnowitt1962}\footnote{The ideas 
were first seen in the long out of print {\it Gravitation: an introduction to 
current research}. The authors have later on released the article on ArXiv as 
cited.}. Hamiltonian formulation has also surfaced in the works
of Ashtekar \cite{ashtekar87}. The first papers on the subject often neglected the boundary terms, 
however, later works have clarified these details ({\it e.g.} \cite{Brown1992, 
Hawking1995}). Hamiltonian formulation has not 
received too much interest in contemporary papers. In particular and 
to our knowledge the use of Hamiltonian formulation on perturbations of $f(R)$ theories
has not been studied so far. The main interest has been in specific 
choices for the function $f(R)$.

In the present paper we look into perturbations using Hamiltonian formalism
of $f(R)$ theories. While the technique has not yet been
applied to general $f(R)$ theories with perturbations it is a useful tool 
in studying the stability of $f(R)$ models:
with it is simple to study perturbations perpendicular 
to $R$. As in classical mechanics the Hamiltonian is written as a
functional of the fields and their canonical momenta. However, in a
geometric theory like General Relativity and $f(R)$ theories, some complications
appear due to constraints between field components. The two main aspects of the canonical Hamiltonian formalism are 
that the field equations are of the first order in the time derivatives 
and that time is distinguished from other coordinates. For writing the
Hamiltonian equations we must thus foliate the region of space-time with 
space-like hypersurfaces. Finally the resulting field equations for the 
perturbations are then analyzed for instabilities. The conventions and 
details of the formalism can be found in \cite{poisson2004}.

The paper is organized as follows. In section \ref{hamform} we write the 
Hamiltonian field equations. The 3+1 decomposition and foliation
of the space-time are also presented. The first order perturbations are 
added to the system in section \ref{1order}. We also take a look at second order 
perturbations in section \ref{2order}. In section \ref{concl} we summarize our results.

\section{Hamiltonian formulation}\label{hamform}

In this section we mainly follow the treatment of \cite{poisson2004}. Another overview 
of Hamiltonian (and Lagrangian) formulation of general relativity can be found in 
\cite{wald1984}. As $f(R)$ theories of gravity can be written as
scalar tensor theories 
\cite{Chiba2003}, we start by writing the $f(R)$ action as 
\ba
S&=S_G+S_M=\inv{16\pi}\int_V f(R)\sqrt {-g}\ d^4x+S_s+S_M  \nonumber\\
&=\inv{16\pi}\int_V \sqrt {-g}\big(f(\varphi)+f'(\varphi)R-
f'(\varphi)\varphi)\big)d^4x+S_s+S_M, \label{frsc}
\end{align}
where $V$ is a volume of space-time, $S_s$ is a surface term
which we will cover later on, $S_M$ is the matter term and $\varphi$
is the auxiliary scalar field. Variation with respect to $\varphi$ would lead to
equation $f''(\varphi)\big(R-\varphi\big)=0$. Therefore $R=\varphi$
unless $f''(\varphi)=0$. If $f''(\varphi)=0$ the two forms of action
(\ref{frsc}) are trivially equal. From here on we make the assumption
$f''(\varphi)=0$ unless stated otherwise. 

For the purposes of writing the action in terms of the Hamiltonian a
3+1 decomposition is needed. We foliate the space-time volume $V$ with
space-like hypersurfaces $\Sigma_t$ of constant time. We shall use the Greek alphabet for space-time and the Latin alphabet for space. By decomposing the line element 
to (three) scalar, (three) vector and (three) tensor parts as
\be
ds^2=-N^2dt^2+h_{ab}(dy^a+N^adt)(dy^b+N^bdt),\label{admline}
\ee
one defines the lapse $N$, the shift $N_a$, and the induced metric
$h_{ab}=g_{\mu\nu}e^\mu_ae^\nu_b$ on the hypersurface $\Sigma_t$. By using these 
introduced quantities the invariant volume element reads $\sqrt {-g} \ d^4x=N\sqrt h \ dtd^3x$. 
The Ricci scalar can be written in terms of the extrinsic curvature
$K_{ab}$, $n_\alpha$ the unit normal to the boundary $\partial V$ and $\tilde R$, 
the induced Ricci scalar of the (three dimensional) metric $h_{ab}$ as
\be
R=\tilde R+K^{ab}K_{ab}-K^2-2\nabla_\alpha(\nabla_\beta n^\alpha n^\beta-
n^\alpha\nabla_\beta n^\beta).
\ee
Extrinsic curvature is the measure of shrinkage and deformation of an object 
upon being moved a unit interval of proper time into the enveloping space-time.
It can be written as a function of the induced metric, the lapse and the shift, which appear to be the fields we are finally
interested in,
\be
K_{ab}=\inv{2N}\big(\dot h_{ab}-N_{a|b}-N_{b|a}\big). \label{kab}
\ee
The surface terms of the action (\ref{frsc}) are not
of special interest in this paper. However, it is not trivial that these parts
do not affect the results. Generally, the surface term must be added to the action 
in order to avoid the need for further boundary conditions ({\it e.g.} 
\cite{querella98}). 

By choosing the space-time volume $V$ so that, its boundary can 
be written as a union of two space-like hypersurfaces $\Sigma_{t_2},\,-\Sigma_{t_1}$ 
with normals pointing outwards and a time-like hypersurface $\mathcal B$, {\it i.e.}
$\partial V=\Sigma_{t_2}\cup(-\Sigma_{t_1})\cup\mathcal B$. The surface term
reads 
\ba
S_s=&\inv{8\pi}\oint_{\partial V}\epsilon f'(\varphi)K|h|^{1/2}d^3y-S_0 \nonumber\\
=&\inv{8\pi}\Big(\int_{\Sigma_{t_1}}f'(\varphi)K\sqrt h \
d^3y-\int_{\Sigma_{t_2}}f'(\varphi)K\sqrt h \ d^3y+\int_{\mathcal B}
f'(\varphi)\mathcal K\sqrt {-\gamma}\ d^3y\Big)-S_0,\label{surft}
\end{align}
where $\epsilon=n^\alpha n_\alpha$. Here 
$S_0=\inv{8\pi}\oint_{\partial V}\epsilon K_0\sqrt{|h|} \ d^3y$ 
is a non-dynamical subtraction term the purpose of which is to
prevent the integral from diverging in the limit when the spatial boundary 
$S_t$ is pushed to the infinity. The constant $K_0$ is the extrinsic curvature of 
the boundary $\partial V$ embedded in flat space-time.  
In the last term $\gamma$ is the induced metric on $\mathcal B$ and 
$\mathcal K$ is the extrinsic curvature scalar of $\mathcal B$.
However, this is not the only term contributing to the surface
part. The term $f'(\varphi)R$ from (\ref{frsc}) produces surface and volume terms, namely
\ba
\int_Vf'(\varphi)R\sqrt
{-g} \ d^4x=&\int^{t_2}_{t_1}dt\int_{\Sigma_t}f'(\varphi)
(\tilde R+K^{ab}K_{ab}-K^2)N\sqrt h \ d^3y \nonumber\\
&-2\oint_{\partial V}f'(\varphi)(\nabla_\beta n^\alpha n^\beta-
n^\alpha\nabla_\beta n^\beta) \ d\Sigma_\alpha.\label{f'}
\end{align}
When combining these surface contributions the first two terms in
\eqref{surft} are eliminated. The only surface term left from \eqref{f'} is
\ba
-2\int_{\mathcal B}f'(\varphi)(\nabla_\beta n^\alpha n^\beta-n^\alpha\nabla_\beta n^\beta)d\Sigma_\alpha
&= -2\int_{\mathcal B}f'(\varphi)(\nabla_\beta n^\alpha) n^\beta r_\alpha\sqrt{-\gamma} \ d^3y\nonumber\\
&= 2\int_{\mathcal B}f'(\varphi)(\nabla_\beta r_\alpha) n^\beta n^\alpha \sqrt{-\gamma} \ d^3y,
\end{align}
where $r_\alpha$ is the perpendicular unit vector of the boundary $S_t$ of 
$\Sigma_t$, {\it i.e.} $r^\alpha r_\alpha=1$ and $r^\alpha n_\alpha=0$. 
Summing the remaining surface terms we obtain
\be
S_S=2\oint_{S_t}(k-k_0)f'(\varphi)N\sqrt{\sigma} \ d^2\theta. \label{ss}
\end{equation} 
We have also introduced the induced metric on the boundary $S_t$, 
$\sigma_{AB}=h_{ab}e^a_Ae^b_B$ and $\sigma$ is its trace. The extrinsic curvature 
of $S_t$ embedded in $\Sigma_t$ is $k_{AB}e^a_Ae^b_B\nabla_br_a$, $k$ is its 
trace and similarly $k_0$ with the embedding in flat space. The constant $k_0$ comes from 
the subtraction term. 

We now have the surface part of the action ready for construction of the Hamiltonian. We shall see later on that the surface term \eqref{ss} is indeed cancelled in the process of calculating the field equations. Many of the technical details were omitted and we refer the reader to \cite{poisson2004}. The generalization to $f(R)$ is easy. 

In the Hamiltonian formulation field equations are found for fields and their momentum conjugates. Here 
the fields are $h_{ab}$, $N$, $N_a$ and $\varphi$. It turns out that in the case of $f'(R)$ gravity we need only the momentum conjugate to the induced 
metric $h_{ab}$. This can be written using the extrinsic curvature 
\be
p^{ab}=\frac{\partial K_{cd}}{\partial \dot h_{ab}}\frac{\partial}{\partial K_{cd}}
\big(\sqrt g\mathcal L_G\big)
=\frac{\sqrt h f'(\varphi)\big(K^{ab}-K h^{ab}\big)}{16\pi}. \label{pab}
\ee
For evaluating $\partial K_{cd}/\partial\dot h_{ab}$ the extrinsic curvature was 
written as a function of the induced metric given in the formula (\ref{kab}).

For writing the Hamiltonian density $\mathcal H=p^{ab}\dot h_{ab}-\sqrt{-g}\mathcal L$ we still need the volume 
part of the action. We write the gravitation part of the action without the surface part (which we include 
later on) as
\ba
S_{GV}=\inv{16\pi}\int_Tdt\Big\{\int_{\Sigma_t}\big[f(\varphi)+f'(\varphi)(K^{ab}K_{ab}+
\tilde R-K^2)-f'(\varphi)\varphi\big]N\sqrt h \ d^3x\Big\}.
\end{align}
After some manipulations the volume part of the Hamiltonian density can be cast to the form
\ba
\mathcal H_G&=p^{ab}\dot h_{ab}-\sqrt {-g} \mathcal L_G \nonumber\\
&=\frac{N\deth}{16\pi}\Big\{f'(\varphi)\big[K^{ab}K_{ab}-K^2-\tilde R+\varphi\big]-f(\varphi)\Big\}\nonumber\\
&\ \ \ +\frac{\deth f'(\varphi)}{8\pi}\big[(K^{ab}-Kh^{ab})N_a\big]_{|b}-\frac{\deth f'(\varphi)}{8\pi}(K^{ab}-Kh^{ab})_{|b}N_a.
\label{hamden}
\end{align}

To express the Hamiltonian density in term of adequate variables, {\it i.e.} induced metric $h_{ab}$ and its conjugate momentum  
$p_{ab}$, we need to rewrite the extrinsic curvature. By inverting \eqref{pab} we get
\be
\deth K^{ab}f'(\varphi)=16\pi(p^{ab}-\half ph^{ab})\equiv\hat p^{ab}-\half \hat ph^{ab}.
\ee
Using this equation the Hamiltonian can be written as a function of the momentum conjugate. 
Now the volume part of the gravitational Hamiltonian is obtained by integrating $\mathcal H_G$ over the hypersurface $\Sigma_t$
\ba
H_G=&\inv{16\pi}\int_{\Sigma_t}\Big\{N\sqrt h \big[\varphi f'(\varphi)-
f(\varphi)-\tilde Rf'(\varphi)\big]+\frac N{\sqrt h f'(\varphi)}\Big(\hat p_{ab}\hat p^{ab}
-\frac{\hat p^2}2\Big)-\\
&-2\sqrt hN_a\Big(\frac{\hat p^{ab}}{\sqrt h}\Big)_{|_b}\Big\}d^3x.
\end{align}
Similarly we get the surface part of the gravitational Hamiltonian by taking the appropriate terms and integrating over the 
hypersurface $\Sigma_t$:
\be
H_S=\inv{8\pi}\oint_{S_t}\Big[N(k-k_0)-\frac{N_a \hat p^{ab}r_b}{\sqrt h}\Big]f'(\varphi)\sqrt\sigma d^2\theta.
\ee
The latter term is produced by applying Gauss theorem to the middle term of \eqref{hamden} when integrating over the density. 
We obtain the field equations by varying the action with respect to $N, N^a$, $h_{ab}, p_{ab}$ 
and $\varphi$. As we can see in the action the only time derivatives are those of the induced metric. Thus the only dynamic field is $h_{ab}$ and the only momentum conjugate needed is $p_{ab}$. We have the normal boundary conditions for the variations vanishing 
on the boundary
\be
\delta N=\delta N^a=\delta h_{ab}=\delta\varphi=0.
\ee
The full Hamiltonian $H$ includes both surface and volume parts as well as a matter part $S_M$. 
Since we can write variation of the action as
\be
\delta S=\int^{t_2}_{t_1}dt\Big[\int_{\Sigma_t}(p^{ab}\delta\dot h_{ab}+\dot h_{ab}\delta p^{ab})d^3y-\delta H\Big]
\ee
the Hamiltonian equations are of the form
\be
\dot h_{ab}=\frac{\partial \mathcal H_G}{\partial p}, \ \dot p^{ab}=-\frac{\partial \mathcal H_G}{\partial h}+\frac{\delta S_M}{\partial h}, \ \frac{\partial \mathcal H_G}{\partial N_a}=0, 
\ \frac{\partial \mathcal H_G}{\partial N}=\frac{\delta S_M}{\partial N}, \ \frac{\partial \mathcal H_G}{\partial\varphi}=0.
\ee
To simplify the field equations, we can choose the foliation to be such that $N_a=0$ 
and hence $h_{ab}=g_{ab}$, when the effects of the surface terms vanish.
This choice removes one field equation, that of $N_a$, and the other ones
are much simplified.  After tedious calculations we end up with equations
\begin{subequations}\label{eqmo}
\ba
-\frac{\dot p^{ab}}{N\deth}&=G^{ab}f'(\varphi)+\frac {h^{ab}}2\Big(\varphi f'(\varphi)-f(\varphi)-16\pi P-\frac{p^{cd} p_{cd}-
\frac{p^2}2}{h f'(\varphi)}\Big)+\frac{2 p^a_c p^{cb}-p \ p^{ab}}{hf'(\varphi)},\label{eqpab}\\
16\pi \sqrt h\rho&=\big(\tilde R+K^2-K^{ab}K_{ab}-\varphi\big)f'(\varphi)\sqrt h+f(\varphi)\sqrt h,\label{eqn}\\
\dot h_{ab}&=\frac{2N}{\sqrt h f'(\varphi)}\big(p_{ab}-\half p h_{ab}
\big),\label{eqp}\\
\varphi-\tilde R&=\frac{\frac{\hat p^2}2-\hat p_{ab}\hat p^{ab}}{h\big(f'(\varphi)\big)^2},\label{phieq}
\end{align}
\end{subequations}
where $\tilde G^{ab}=\tilde R^{ab}-\frac 12\tilde Rh^{ab}$. 
For technical details we refer the reader to \cite{poisson2004} which can straightforwardly be generalized to the 
$f(R)$ case. Note, however, that in the derivation of the equation 
\eqref{phieq} further use is made of the assumption $f''(R)\neq 0$. Otherwise, we would get a trivial equality.
As can be seen in equations \eqref{eqpab} 
and \eqref{eqn} we have also added matter
\ba
\frac{\delta S_m}{\delta N}&=-\frac{\sqrt g}2 T_{00}\frac{\delta g^{00}}{\delta N}=-\deth\rho, \\
\frac{\delta S_m}{\delta h_{ab}}&=-\frac{\sqrt g}2 T^{ab}\frac{\delta g_{ab}}{\delta h_{ab}}=-\frac{N\deth}2 Ph^{ab},
\end{align}
which is of the perfect fluid form.

Even though we assumed from the start that $f''(R)\neq 0$ it is worthwhile to take look at the case of 
Einstein-Hilbert Lagrangian. If in Eq. \eqref{frsc} we choose $f(R)=R$, the equality is trivial, and only a variation of a constant resulting in a trivial field equation. As the assumption of $f''(R)\neq 0$ is needed 
in the field equations only in \eqref{phieq} the equations would stay the same except for this one equation 
which is irrelevant. From equation \eqref{eqn} we get the familiar Friedmann equation for the background
\be
\mathcal H^2 =\frac{8\pi\rho_0}{3a}\label{rfried},
\ee
in a matter dominated universe ($\rho=a^{-3}\rho_0$). Here $\mathcal H=a'(\eta)/a(\eta)$ is the conformal Hubble parameter. We will need this background result later on when we insert the asymptotic background solution into the equations. Namely, we can solve $a'(\eta)$ from this equation.

\section{First order perturbations}\label{1order}

In this section we add first order perturbations to the metric and the momentum conjugate. 
%We do not introduce perturbations to matter as they are related to the curvature scalar. 
In general relativity the trace equation connects curvature and matter density (for fixed equation of state
$p=p(\rho )$)
%for a matter filled universe ($w=0$) 
by a simple relation $R=\kappa(\rho-3P)$, where $\kappa = 8\pi G$. The perturbations would be 
connected correspondingly: $\delta R=\kappa(\delta\rho-3\delta P)$. As the trace equation in $f(R)$ gravity is 
$f'(R)R-2 f(R)+3\Box f'(R)=\kappa(\rho-3P)$ there are more freedom in 
metrics that produce a given mass configuration. Indeed, the relation between curvature and matter distribution is 
no more an algebraic one, but defined by an differential equation. Thus the phase space of metrics is larger and 
there are perturbations keeping $R$ and thus $\rho$ fixed.
This is manifested by the statement that Birkhoff's theorem\footnote{Birkhoff proved the so called Birkhoff theorem in 
1923 \cite{birkhoff}. However, two years earlier a less known Norwegian physicist Jebsen presented the idea in \cite{jebsen}. The history of the theorem is examined in \cite{Johansen2005}.} is no more valid in the traditional form in $f(R)$ theories \cite{Bronnikov1994,Schmidt1997}.  
Since there are number of studies of the perturbations along $R$ ({\it e.g.} \cite{Dolgov2003}) we are now interested 
in the opposite and do not introduce perturbations to matter but perturbations perpendicular to $R$ only,
{\it i.e.} $\delta R=\delta \rho =0$.

We may add the most general first order perturbations to the metric. These include scalar, vector and tensor 
perturbations. In light of the recent observations and for simplicity we examine the case of spatially flat FRW metric. 
The perturbations in first order can now be written as \cite{Bartolo2004}
\begin{subequations}\label{fometgeng}
\ba
g_{00}&=\tilde g_{00}-2a^2\Phi, \\
g_{0a}&=\tilde g_{0a}+a^2(\partial_a\omega+\omega_a), \\
g_{ab}&=\tilde g_{ab}+a^2\big(-2\Psi\delta_{ab}+\nabla_{ab}\chi+\partial_a\chi_b+\partial_b\chi_a+\chi_{ab}\big),
\end{align}
\end{subequations}
where tilde denotes the background part and the vectors $\omega^a$ and $\chi^a$ are transverse ({\it i.e.} $\partial^a\omega_a=0, \ \partial^a\chi_a=0$) and $\chi_{ab}$ is trace free and symmetric tensor ({\it i.e.} $\partial^a\chi_{ab}=0, \ \chi^a_a=0$). Comparing the elements in \eqref{fometgeng} and the line element \eqref{admline} to find the perturbations in the first order for 
lapse, shift and the induced metric we obtain
\begin{subequations}\label{fometgen}
\ba
N&=\tilde N+a\Phi, \\
N_a&=\tilde N_a+a^2(\partial_a\omega+\omega_a)\equiv \tilde N_a+a^2\hat\omega_a, \\
h_{ab}&=\tilde h_{ab}+a^2\big(-2\Psi\delta_{ab}+\nabla_a\nabla_b\chi+\partial_a\chi_b+\partial_b\chi_a+\chi_{ab}\big) \nonumber\\
&\equiv\tilde h_{ab}+a^2\big(-2\Psi\delta_{ab}+\hat\chi_{ab}\big).
\end{align}
\end{subequations}

The standard practice of splitting the perturbations into scalar, vector and tensor parts \cite{Bertschinger1993} is 
motivated by the reason that in a linear theory these modes decouple. Moreover each of them has a clear 
physical interpretation \cite{mukhanov1990}. The first order vector perturbations are not generated in the presence 
of scalar perturbations and dissipate over time. Tensor perturbations cause gravitational waves which do not couple to 
first order scalar perturbations. Therefore, we may omit vector and tensor perturbations in the first order case and 
assume 
$\omega_a=0, \ \chi_{ab}=0$. We can further simplify the metric for our purposes by choosing an appropriate gauge. We 
choose to use the Poisson gauge \cite{Bartolo2004} which is a generalization of the much used longitudinal gauge. The 
gauge conditions are 
\begin{subequations}\label{poissong}
\ba
\nabla\cdot\mathbf{\hat\chi} &=0, \\
\nabla\cdot\mathbf{\hat\omega} &=0.
\end{align}
\end{subequations}
Since $\omega^a$ and $\chi_a$ are transverse vectors and $\chi_{ab}$ is a symmetric, transverse and trace-free tensor 
we have $\omega=\chi=\chi_a=0$. Along with the physical meaning of the perturbations discussed above the 
perturbed metric simplifies to
\begin{subequations}\label{fomet}
\ba
N&=\tilde N+a\Phi, \\
N_a&=\tilde N_a, \\
h_{ab}&=\tilde h_{ab}-2a^2\Psi\delta_{ab}.
\end{align}
\end{subequations}

%We have now introduced perturbations to the metric. 
As the dynamical components of the metric are coupled to their momentum conjugates, 
we are to add perturbations also to the conjugates. 
%Otherwise the perturbations would be strictly limited. 
Only the induced metric $h_{ab}$ has a conjugate $p_{ab}$, and hence for perturbed one we write  
\be
p_{ab}=\tilde p_{ab}+\Theta\delta_{ab}\label{fomomp}
\ee
having same structure as (\ref{fomet}c).

In the following we work mostly, unless otherwise stated, in conformal time instead of 
standard coordinate time. So we have $ds^2=-a(\eta)^2d\eta^2+a(\eta)^2\delta_{ab}dx^adx^b$, where the conformal time 
$\eta$ is related to  standard coordinate time by $d\eta=a^{-1}dt$. Prime denotes derivatives with respect to 
the conformal time and dot denotes derivatives with respect to the 
coordinate time. This choice of background metric corresponds to $\tilde R=\tilde G_{ab}=0$ and $\sqrt{\tilde h}=a^3$. Also, we now have 
$\tilde p_{ab}=-2f'(\varphi)a^3a'$. Since we wrote the $f(R)$ theory using a scalar in \eqref{frsc} we have $\varphi\sim R$. Perturbing $\varphi$ would produce perturbations parallel to $R$ which we are not interested in. 

The Eqs. \eqref{eqmo} for the chosen background metric and scalar field are now given in a fairly simple form. This reads
\begin{subequations}
\ba
16\pi P a^4&=2(a')^2f'(\varphi)-a^4\big(f(\varphi)-\varphi f'(\varphi)\big)-4a\big(a''f'(\varphi)+
a'\varphi'f''(\varphi)\big), \label{bgp}\\
16\pi\rho a^3&=\frac{f'(\varphi)\Big(6(a')^2-a^4\varphi\big)}a+a^3f(\varphi),\label{bgh} \\
\varphi&=\frac{6(a')^2}{a^4}\label{scalar}.
\end{align}
\end{subequations}
We get only three non-trivial equations as \eqref{eqp} produces only a trivial identity. These equations, satisfied for any acceptable matter are used to simplify the perturbation equations derived later. In the following we assume a matter filled universe with $P=0$ and $\rho=\rho_0/a^3$.

By adding the perturbations introduced in \eqref{fometgen} and \eqref{fomomp} to the equations of 
motion \eqref{eqmo} we get three equations for the large scale perturbations ({\it i.e.} space independent
perturbations)
\begin{subequations}\label{pereqmo}
\ba
\Psi&=\frac{\Theta}{10a^3a'f'(\varphi)}, \label{psieq}\\
\Theta'&=\Big(\frac{3a'}a+\frac{a''}a
+\frac{12a'f''(\varphi)\big(aa''-2(a')^2\big)}{a^5f'(\varphi)}\Big)\Theta, \label{thetaeq}\\
\Phi&=0,
\end{align}
\end{subequations}
where we in \eqref{thetaeq} the background equations \eqref{bgp} and \eqref{scalar} were applied 
to simplify the equation. We immediately notice, that there remains only one dynamic equation while the other two are algebraic.
The background equation for the induced metric can be used to eliminate the second time derivative of the scale parameter. Eq. (\ref{thetaeq}) is thus written as 
\be
\Theta'=\Big(\frac{5a'}a-\frac{4\pi\rho_0}{a'f'(\varphi)}\Big)\Theta. \label{thetae}
\ee
The behaviour of perturbation is clearly dependent on form of the function $f(\varphi)$ explicitly
via its derivatives. Moreover, it is found that the time derivative of $\Psi$ is zero and therefore 
by equation \eqref{psieq} we can write
\be
\Theta=C a^3a'f'(\varphi)
\ee
where $C$ is a constant. So, in a universe with growing $a(\eta)$ the 
perturbations in momentum conjugate increase. The perturbations of metric tensor,
however, behave differently: the temporal part vanishes and the spatial perturbations are constant. 
So, the system leaves the linear perturbative regime and ultimately suffers linear instability.

Although asymptotic analysis is ultimately irrelevant for a linearized unstable system, we take a look 
to some examples to get a better feeling of the evolution. As known, the simple function 
$f(R)=R-\mu^4/R$ results asymptotic Einstein-de-Sitter behaviour. Now $a(t)=e^{\Lambda t}$, and coordinate and 
conformal times are related by $\eta=-\frac{e^{-\Lambda t}}{\Lambda}+c$ so that $a(\eta)=\Lambda^{-1}(c-\eta)^{-1}$. 
In the high curvature limit we get
\[
\Theta(\eta)=\hat C\frac{36\Lambda^4+\mu^4}{36(c-\eta)^5\Lambda^8}
\]
where $\hat C$ is a constant. The result can also be written more intuitively in coordinate time as
\[
\Theta(t)=C e^{4\Lambda t}\big(1+\frac{e^{4\Lambda t}\mu^4}{36\Lambda^4}\big),
\]
and thus the perturbations increase as time goes to infinity. Here $C$ is another constant. Ultimately the first order perturbation theory 
breaks down; it is not applicable in this case. Similar behaviour can seen explicitly for another often used
$f(R)=R-\mu^2R^2$. 

Even though we have not included perturbations in matter it is worthwhile to check what would happen if we 
did include these perturbations. For a moment we consider  $\rho=\tilde\rho+\sigma$, where $\sigma$ is a 
perturbation. It turns out that no density perturbations are present, {\it i.e.}
perturbation equation is $\sigma=0$. This is not surprising as the matter perturbations are coupled to the 
temporal perturbation of the metric which is also zero. These vanish unless $\varphi$ (which is essentially $R$)
is perturbed, too. 

As we have found, the only dynamical equation is \eqref{thetaeq} for the momentum conjugate,
while the two other equations determine, how metric perturbations follow it; they are
constraint equations. If these constraints were to be discarded, we would end up with non-diagonal 
perturbations in the metric. Moreover non-existence of temporal perturbations is connected with the orthogonality 
of perturbations to curvature. As it appears that the spatial perturbations in the metric do not grow or vanish
in time, there is a flat direction of phase space, where any spatial first order perturbation is possible and stable.

\section{Second order perturbations}\label{2order}

We have now seen that the first order perturbation predicts that $f(R)$ theories suffer instability which
invalidates first order expansion; equation \eqref{thetae} reveals that we cannot use first order perturbation 
theory. The next check would be to consider second order perturbations, which might give us further understanding 
of the perturbations involved. We first write the most general form of the metric and the conjugate momentum as
\begin{subequations}\label{sometgen}
\ba
N&=\tilde N+a(\Phi^{(1)}+\Phi^{(2)}), \\
N^a&=\tilde N^a+a\sum^2_{r=1}(\partial_a\omega^{(r)}+\omega_a^{(r)}), \\
h_{ab}&=\tilde h_{ab}+a^2\Big[-\big(2\Psi^{(1)}+\Psi^{(2)}\big)\delta_{ab}+\sum^2_{r=1}\big(\nabla_a\nabla_b\chi^{(r)}
+\partial_a\chi^{(r)}_b+\partial_b\chi^{(r)}_a+\chi^{(r)}_{ab}\big)\Big], \\
p_{ab}&=\tilde p_{ab}+(\Theta^{(1)}+\Theta^{(2)})\delta_{ab},
\end{align}
\end{subequations}
where the upper index $i$ denotes the order of the perturbation. As we have chosen to work in the 
Poisson gauge \cite{Bertschinger1993} we have $\omega^{(r)}=\chi^{(r)}=\chi_a^{(r)}=0$. The 
vector perturbations $\omega_a^{(r)}$ and $\chi_{ab}^{(r)}$ still remain, however, and some extra attention
has to be paid to them. In general the scalar, vector and tensor perturbations do not decouple any more in 
the second order perturbation theory. 
First order vector perturbations contribute to the second order scalar perturbations by terms like 
$\omega^a\omega_a$ and vice versa. However, first order perturbations do not manifest themselves if not 
present initially. Since we are now interested in to show the instability of the system,
it is sufficient that some initial condition reveals unstable behaviour. In particular we are free to
choose initial condition $\omega_a(0)=0$ for the vector perturbations. First order tensor 
perturbations  can omit them as well. Note, that if we were trying to show
the stability of the system, the burden of proof would be much heavier: we should show, that any choice of
initial conditions leads to stable system. 

As mentioned, vector and tensor perturbations in second order cannot be discarded by similar arguments.
They are strongly affected by first order scalar perturbations. However, the second order scalar perturbations 
are again independent of the tensor and scalar perturbations of the second order. Therefore, for our purposes, 
it is sufficient to study only second order scalar perturbations, which can be performed rather simply.
We write the relevant perturbation equations for second order in the same manner as in the 
previous section. We obtain 
\begin{subequations}\label{ndequations2}
\ba
{\Psi^{(2)}}'&=0, \\
\Theta^{(2)}&=-\frac{3}{20a^3a'f'(\varphi)}\big(\Theta^{(1)}\big)^2+5a^3a'f'(\varphi)\Psi^{(2)}, \\
\Phi^{(2)}&=0.
\end{align}
\end{subequations}
Thus metric perturbation $\Phi^{(2)}$ still vanishes and $\Psi^{(2)}$ is again constant related to
the perturbation of the momentum conjugate $\Theta^{(2)}$ by (\ref{ndequations2}b).
The perturbation in the momentum conjugate is still depending on the form of the $f(R)$. For 
$f(R)=R-\mu^4/R$ the result is the same as in the first order; the perturbation of the temporal part disappear, the spatial part remains constant and the momentum conjugate is has the only dynamical equation. It is clear that the 
instabilities in the first order propagates to the second order as the metric perturbations  behave 
in exactly the same way in both first and second order. Thus $f(R)$ models may be inherently unstable up to
second order when examining perturbations perpendicular to $R$. Because of the similar form of the first and the
second order scalar perturbations one might conjecture that it is a more general feature of the theory.

\section{Discussion}\label{concl}

Traditionally the stability analysis is performed in the Lagrangian formalism and
the analysis parallel to $R$ has been carried out before in several papers ({\it e.g.} \cite{Dolgov2003}).
Many of the interesting $f(R)$ models have been found to be inherently unstable in the past \cite{Kainulainen20071,Faraoni2005}. 
However, stability analysis has not yet been used to the full extent as long as the studies concentrate on 
curvature perturbations only. By using the Hamiltonian instead of Lagrangian formulation we examined the large 
scale cosmological perturbations perpendicular to $R$ with non-relativistic matter. These perturbations are 
fairly easy to examine with the Hamiltonian formulation. The found instabilities are noticeably different to 
those of previous works 
({\it e.g.} \cite{Faraoni2005}). Because of the constraint $\dot R=0$ diagonal perturbations of
metric and conjugate momentum are related to each other. The temporal part of the metric showed up to be 
constrained by the conjugate momentum one. Moreover, the spatial part of the metric is forced to vanish. 
If these constraints were not satisfied we would have non-trivial perturbations of non-diagonal elements of
the metrics like $g_{0i}$.

The perturbations of the momentum conjugate turn out to be the most interesting ones. The equation depends 
explicitly on the form of the function $f(R)$. Some choices of $f(R)$ lead clearly unstable cosmological model,
but seemingly not all. We have studied some well-known $f(R)$ functions and find them unstable.
Albeit the physical interpretation of the perturbation momentum 
conjugate is unfortunately not as clear as that of the metric perturbations, equation \eqref{pab} demonstrates
the relation between the momentum conjugate and the extrinsic curvature. In the 3+1 decomposition the intrinsic 
curvature $\tilde R$ defines how the hypersurface is curved whereas the extrinsic curvature defines how each slice 
is curved relative to the enveloping space-time.

As the perturbations were not well-behaved in this context further studies would be relevant in order to find 
the limits of these constraints. Fruitful directions would likely to be investigating the effects of more other 
types of matter. Also, it would be prudent to examine the case where the metric can include shift ({\it i.e.} $N_a\neq 0$). 
It is clear from the form of \eqref{hamden} that such a generalization would affect the following equations of 
motion deeply as the last term would be non-zero. This is understandable as the metric would now include spatio-temporal elements. It is also possible to study more general theories with the Lagrangian depending also on for 
example $R_{\mu\nu}R^{\mu\nu}$ or Gauss-Bonnet term. 

It appears that with Hamiltonian formulation of perturbations can be used to constrain the spectrum of 
cosmologically acceptable $f(R)$ theories. While there are several physical arguments to judge the $f(R)$
theories like cosmological observations and solar system behaviour, stability analysis is one important
tool to rule out ill-behaved models out of numerous possible modified theories of gravity. With 
continued studies it is possible to find the ones best describing the observed behaviour of the universe.  

\subsection*{Acknowledgments}

This project has been partly funded by the Academy of Finland project no. 8111953. JV is also supported by the Magnus Ehnroothin s\"a\"ati\"o foundation.

\addcontentsline{toc}{chapter}{Viitteet}
\bibliography{refs}

\end{document}